\documentclass{article}
\usepackage[dvips]{graphicx,epsfig}
\begin{document}
\hrule
\vskip4mm
\begin{center}
{\bf
Proceedings of the Workshop {\it SZ Toulouse} \\
Observatory of Midi-Pyr\'en\'ees,\\
 Toulouse (France), June 29-30th, 2000\footnote{Talk given by D. Puy 
(puy@physik.unizh.ch)}}
\end{center}
\vskip4mm
\hrule
\vskip8mm
\begin{center}
\huge{\bf Shape and geometry of galaxy clusters and the SZ effect}
\vskip3mm 
\Large{\bf (Determination of the Hubble constant)}
\vskip8mm
\large{\bf L. Grenacher$^{1,2}$, 
Ph. Jetzer $^{2,3}$, R. Piffaretti$^{1,2}$, D. Puy$^{1,2}$, 
M. Signore$^4$}
\vskip2mm
$^1$ Paul Scherrer Institute, LAP, Villigen (Switzerland)
\\
$^2$ Institute of Theoretical Physics, Univ. Z\"urich (Switzerland)
\\
$^3$ Institute of Theoretical Physics, ETH Z\"urich (Switzerland)
\\
$^4$ Observatoire de Paris, DEMIRM (France)
\end{center}
\begin{abstract}
We discuss the influence of the finite extension and the geometry of clusters of galaxies, as well as of the polytropic temperature profile, on the determination of the Hubble constant.
\end{abstract}
\section{Introduction}
The SZ effect (Sunyaev-Zel'dovich 1972) offers the possibility to put 
important constraints on the cosmological models. Combining the temperature 
change in the cosmic microwave background due to the
SZ effect and the X-ray emission observations, 
the angular distance to galaxy clusters, and consequently the Hubble constant $H_o$, can be derived.\\  
The SZ effect is difficult to measure, since systematic errors 
can be important.  For example, Inagaki et al. (1995) 
analysed the reliability of the Hubble constant measurement based on 
the SZ effect. Cooray (1998) showed that projection effects of clusters 
can lead incidence on the calculations of the Hubble constant and the gas 
mass fraction, and Hughes \& Birkinshaw (1998) as well as Sulkanen (1999) 
pointed out, that galaxy cluster shapes can produce systematic errors 
on the measured value of $H_o$.
\\
The aim of this contribution is to investigate the influence of extension, shape and temperature profile of the cluster gas distribution on the inferred 
value of $H_o$. In Section 2 we recall the calculations for the 
determination of the Hubble constant for a non-spherical geometry and finite extension of the galaxy cluster, and in Section 3 we present a quantitative discussion on the incidence of these effects on the value of the Hubble constant. In Section 4 we give a short outlook.

\section{Basic equations}
The $\beta$-model (Cavaliere \& Fusco-Femiano 1976) is 
widely used in X-ray astronomy to 
parametrise the gas density profile in clusters of galaxies by fitting their 
surface brightness profile. Nevertheless, fitting an aspherical distribution 
with a spherical $\beta$-model can lead to an important inaccuracy  
(see Inagaki et al. 1995).\\
 Fabricant et al. (1984) showed a pronounced ellipticity for the cluster 
Abell 2256, indicating that the underlying density profile has to be 
aspherical. Allen et al. (1993) obtained the 
same conclusion for the profile of Abell 478, Hughes et al. (1988) for the 
Coma cluster, Neumann \& B\"ohringer (1997) and Hughes \& Birkinshaw 
(1998) for CL0016+16.
\\
Given these observations, we assume an ellipsoidal $\beta$-model
\footnote{The set of coordinates $r_x$, $r_y$ and $r_z$, as well as the 
characteristic lengths of the half axes of the ellipsoid $\zeta_1$, $\zeta_2$ 
and $\zeta_3$ are defined
in units of the core radius $r_c$.}:
\begin{equation}
n_e(r_x, r_y, r_z) = 
n_{eo} \left[ 1 + \frac{r_x^2}{\zeta_1^2} + \frac{r_y^2}{\zeta_2^2}
+ \frac{r_z^2}{\zeta_3^2} \right ]^{-3 \beta /2 }~,\label{eq:dp}
\end{equation}
where $n_{eo}$ is the electron number density at the center of the cluster and 
$\beta$ is a free fitting parameter which lies in the range $1/2 
\leq \beta \leq 1$.
\\
The Compton parameter $y$ and the X-ray surface brightness $S_X$ depend on 
the temperature of the hot gas $T_e$ and the electron number density $n_e$ 
as follows  
\begin{equation}
y  \propto  2  \, \int_{0}^{l} \, n_e T_e dr_y~,
\end{equation}
\begin{equation}
S_x  \propto  2  \, \int_{0}^{l} \, n_e^2 \, \sqrt{T_e} dr_y~,
\end{equation}
where $l$ is the maximal extension of the hot gas along the line of sight 
in units of the core radius $r_c$ and the X-ray emissivity is assumed to be $\epsilon_X=\epsilon\sqrt{T_e}$.
We have chosen the line of sight along the $r_y$ axis.\\
For a detailed calculation of the Compton parameter and the X-ray surface 
brightness we refer to the paper by Puy et al. (2000):  
\begin{eqnarray}
y(r_x,r_z) &=& 
\frac{\kappa_B T_{eo} \sigma_T n_{eo} \zeta_2 r_c}{m_e c^2}
 \times   \left( 1+\frac{r_x^2}{\zeta_1^2} + \frac{r_z^2}{\zeta_3^2} 
\right)^{-\frac{3}{2}\beta + \frac{1}{2}} \nonumber \\
&\times& 
\left[ B\left(\frac{3}{2}\beta-\frac{1}{2},\frac{1}{2}\right) -B_m\left(\frac{3}{2}\beta-
\frac{1}{2},
\frac{1}{2}\right)\right],
\end{eqnarray}
\begin{eqnarray}
S_X(r_x,r_z)&=& 
\frac{\epsilon n_{eo}^2 \sqrt{T_{eo}} \zeta_2 r_c}{4 \pi (1+z)^3} \times 
\left( 1+\frac{r_x^2}{\zeta_1^2} + \frac{r_z^2}{\zeta_3^2} 
\right)^{-3\beta + \frac{1}{2}} \nonumber \\
&\times& 
\, \left[ B\left(3\beta-\frac{1}{2},\frac{1}{2}\right) 
-B_m\left(3\beta-\frac{1}{2},
\frac{1}{2}\right)\right]~,
\label{eq:x}
\end{eqnarray}
where we introduced the Beta and the incomplete Beta-functions 
with the cut-off parameter $m$ given by:
\begin{equation}
m =  \frac{1+(r_x/\zeta_1)^2 + (r_z/\zeta_3)^2}
{1+(r_x/\zeta_1)^2 + (r_z/\zeta_3)^2 + (l/\zeta_2)^2}~.
\end{equation} 
Introducing the angular core radius $\theta_c=r_c/D_A$, where $D_A$ is 
the angular diameter distance of the cluster:
\begin{equation}
D_A \, = \, \frac{c}{H_o} \frac{q_o z + (q_o -1)
(\sqrt{1 + 2q_o z}-1)}{q_o^2 (1+z)^2}~, 
\end{equation}
and $q_o$ is the deceleration parameter,
we can estimate the Hubble constant from the ratio between $y^2(r_x,r_z)$ 
and $S_X (r_x,r_z)$.
If we choose the line of sight through the cluster center we get:
\begin{equation}
H_0(l) =  
 \lambda' T_{eo}^{3/2} \, \frac{S_X(l)}
{y^2(l)} \, \theta_c \,  
\frac{\left[ B\left(\frac{3}{2}\beta-\frac{1}{2},
\frac{1}{2}\right) -B_m\left(\frac{3}{2}\beta-
\frac{1}{2},
\frac{1}{2}\right)\right]^2}{\left[ B\left(3\beta-
\frac{1}{2},\frac{1}{2}\right) -B_m\left(3\beta-\frac{1}{2},
\frac{1}{2}\right)\right]},
\end{equation}
for a finite extension $l$ and, for an infinitely extended cluster, we get instead  
\begin{equation}
H_0 (\infty) =  \lambda' \, T_{eo}^{3/2} \, \frac{S_X(\infty)}
{y^2(\infty)} \, \theta_c \, 
\frac{\left[B(\frac{3}{2}\beta 
-\frac{1}{2}, \frac{1}{2} ) \right]^2}
{ B(3 \beta -\frac{1}{2}, \frac{1}{2} )},
\end{equation}
where $\lambda'$ is a constant (see Puy et al. 2000).\\
Since $S_X$ and $y^2$ are observed quantities, the ratios $S_X(\infty)/y^2(\infty)$ and 
$S_X(l)/y^2(l)$ are in the following both set equal to the measured value 
$(S_X/y^2)_{obs}$.

\section{Determination of the Hubble constant}
\label{SZ:S3}
Recently, Mauskopf et al. (2000) determined the 
Hubble constant from measurements of the X-ray 
emission and millimeter wavelength observations of the SZ effect in the cluster Abell 1835 with the 
Sunyaev-Zel'dovich Infrared Experiment (SuZIE) multifrequency array receiver. 
Assuming a spherical gas distribution with an isothermal equation of state, 
characterised by $\beta=0.58 \pm 0.02$, $T_{eo}=9.8 ^{+2.3}_{-1.3}$ keV and 
$n_{eo} = 5.64 ^{+1.61}_{-1.02} \times 10^{-2}$ cm$^{-3}$, they found a value 
of $H_o^{obs}=59^{+38}_{-28}$ km s$^{-1}$ Mpc$^{-1}$ for the Hubble constant. 
\\
If we suppose other physical characteristics for the cluster such as: 
finite extension, polytropic temperature profile or aspherical 
density distribution, we get of course different values for the $y$-parameter and the surface brightness, and so a relative error with respect to the {\it classical configuration} (i.e. 
spherical distribution with infinite extension and isothermal 
temperature). Thus, we define three kind of relative errors:
\begin{itemize}
\item $\epsilon^{ext}_y \, = \, 1 - ( y_l / y_\infty)$ 
where $y_\infty$ is the Compton parameter for an infinite extension and $y_l$ for a finite cluster extension $l$. Here we consider an isothermal profile and a spherical distribution.
\item $\epsilon^{poly}_y \, = \, 1 - ( y_{poly} / y_{iso})$ 
where $y_{iso}$ is for an isothermal profile and $y_{poly}$ is for a polytropic profile. We consider here a spherical 
distribution with infinite extension.
\item$\epsilon^{geom}_y \, = \, 1 - ( y_{ell} / y_{sph})$  
where $y_{sph}$ is the Compton parameter for a spherical 
distribution and $y_{ell}$ is for an ellipsoidal distribution, both with infinite extension and isothermal temperature profile.
\end{itemize}
Similarly, we can define the relative error for 
the surface brightness $S_X$. 
In the following we discuss the influence of the finite extension and the  
temperature and density profiles on the Hubble constant, and 
compare the result with the value given by Mauskopf et al. (2000).
\subsection{Finite cluster extension}
Since the hot gas in a cluster has a finite extension, each 
of the observed quantities, the Compton parameter and the X-ray surface brightness, will be smaller than those estimated assuming $l \rightarrow \infty$.\\
In Puy et al. (2000) we have analysed the influence of this correction for 
the simplest cluster case: isothermal $\beta=2/3$-model with a spherical 
density profile (i.e. $\zeta_1=\zeta_2=\zeta_3=1$), and a line of sight 
going through the cluster center (i.e. $r_x=r_z=0$). In Table 
\ref{tab:exten} we give the relative error on the Compton $y$-parameter and the surface brightness for different finite extensions of the cluster. For a cluster with an extension of about 10 times the core radius $r_c$, the relative error with respect to the assumption of an infinite extension is only about 7 \% for the Compton parameter. For the X-ray brightness the relative error due to the finite extension is much smaller, for instance an error of about 4\% is obtained, if the cluster has an extension of only 2 times $r_c$.

\begin{table*}[h!tbp]
 \renewcommand{\arraystretch}{1.0}
  \centering
    \begin{tabular}{|c||c|c|c|c|c|}\hline
    $r_c$& 2 &  4 & 6 & 8 & 10 \\ \hline
$\epsilon _{y}^{ext}$ (in \%)&
29 & 15 & 12 & 9 & 7 \\ \hline
$\epsilon _{S_X}^{ext}$ (in \%) &
4 & 1 & 0.4 & 0.2 & 0.1  \\ \hline

    \end{tabular}
    \caption{\small Relative errors $\epsilon_y^{ext}$ on the 
Compton $y$-parameter and $\epsilon_{S_X}^{ext}$ on the surface 
brightness assuming $\beta=2/3$ and a spherical cluster. The line of sight is 
taken to go through the cluster center.}
    \label{tab:exten}
\end{table*}

\begin{figure}[h]
\begin{center}
\epsfig{file=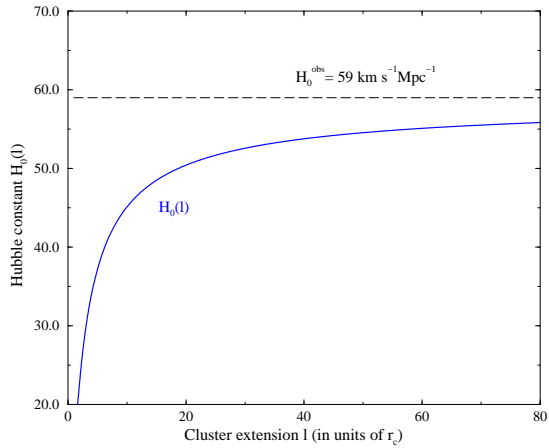,width=6cm,angle=-90}
\caption{\small The Hubble constant $H_o^{obs}$ derived from the data of Mauskopf 
et al. (2000). The curve $H_o(l)$ shows the influence of finite extension. 
The line of sight goes through the center of a spherical, isothermal modelled cluster.}
\label{fig:obs1}
\end{center}
\end{figure}

In Figure \ref{fig:obs1} we show the influence of the finite extension $l$ 
using the same input parameters of Mauskopf et al. (2000). For a spherical 
geometry $H_o$ displays a strong dependence on the cluster extension. 
An extension of $l \sim 10 \, r_c$ leads to $H_o \sim 45$ km s$^{-1}$ Mpc$^{-1}$, which is well below the value found by Mauskopf et al. (2000).

\subsection{Polytropic index}
Recently, Grego, Carlstrom, Joy et al. (2000) observed in Abell 370 a slow 
decline of the temperature with radius, well described by a gas with a 
polytropic index of $\gamma=1.2$. A non-isothermal equation of state for the 
intracluster gas, given by a polytropic temperature profile

\begin{equation}
T_e = T_{eo} \, \left[ \frac{n_{e}}{n_{eo}}\right]^{\gamma-1}~, 
\end{equation}
can lead to a substantial deviation of the estimated quantities when compared to
the isothermal case ($\gamma = 1$). In Table \ref{tab:poly} we have 
summarised the relative errors on $y$ and $S_X$ for different polytropic 
indices. We see that the error can, in some cases, be quite important (i.e. $>10$\%).

\begin{table*}[h!tbp]
 \renewcommand{\arraystretch}{1.0}
  \centering
    \begin{tabular}{|c||c|c|c|c|c|c|} \hline
 polytropic index $\gamma$ &1 & 1.2 & 1.4 & 1.6 & 1.8 & 2 \\ \hline
$\epsilon _{y}^{poly}$ (in \%)&
0 & 20.5 & 30 & 40 & 45 & 50 \\ \hline
$\epsilon _{S_X}^{poly}$ (in \%) &
0 & 4 & 6 & 10 & 12.5 & 15  \\ \hline

    \end{tabular}
    \caption{\small Relative errors $\epsilon_y^{poly}$ on the Compton $y$-parameter and 
$\epsilon_{S_X}^{poly}$ on the surface 
brightness between polytropic and isothermal profiles. The line of sight is 
taken to go through the center of the cluster, which is assumed to have a 
spherical $\beta=2/3$-profile and infinite extension.}
    \label{tab:poly}
\end{table*}

\begin{figure}[h]
\begin{center}
\epsfig{file=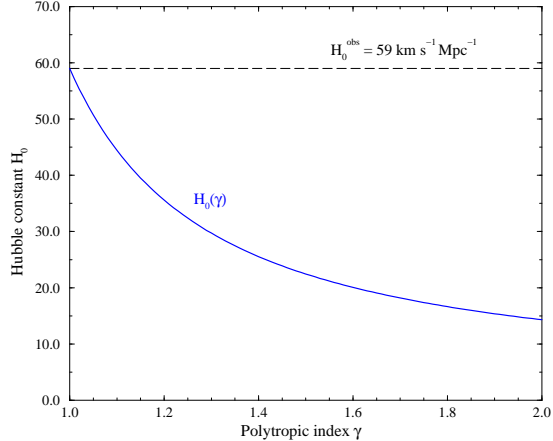,width=6cm,angle=-90}
\caption{\small The curve $H_o(\gamma)$ shows the influence of the 
polytropic index on the Hubble constant, where $H_o^{obs}$ is the Hubble 
constant derived from the data of Mauskopf et al. (2000). The line of 
sight is taken to go through the center of the cluster, which is assumed 
to have a spherical profile and infinite extension.}
\label{fig:obs2}
\end{center}
\end{figure}

In Figure \ref{fig:obs2} we compare the Hubble constant inferred from 
a polytropic temperature profile with the value obtained by Mauskopf et al. 
(2000) for an isothermal profile. We assume a spherical profile with 
infinite extension and obtain a Hubble constant of about 35 instead of 59 km 
s$^{-1}$ Mpc$^{-1}$ taking, as an illustration, a polytropic index of 1.2, as 
estimated by Grego et al. (2000) for Abell 370.

\subsection{Geometrical effect}

Pierre et al. (1996) studied the rich lensing cluster Abell 2390 with ROSAT 
and determined its gas and matter content. They found that on large scales the 
X-ray distribution has an elliptical shape with an axes ratio of minor to 
major half axis of $\zeta_1/\zeta_3\sim 1.33$. 
\\
The influence of the geometrical shape of the cluster profile on the 
investigated quantities are summarised in Table \ref{tab:geom}. 
We considered two axisymmetric cases
{\em prolate} (cigar shaped) with symmetry axis $r_x$, thus 
$\zeta_2=\zeta_3=\sqrt{1/\zeta_1}$, and
{\em oblate} (pancake shaped), with symmetry axis along $r_z$,
and thus $\zeta_2=\zeta_1$ and $\zeta_3=1/\zeta_1^2$. Using our results we 
see that the axes-ratio value obtained by Pierre et al. (1996) leads to a 
relative error on the Compton $y$ parameter of about 10\%, depending on the line of 
sight and the shape of the cluster. The 
surface brightness measurements lead to errors of up to
25\% (see Table \ref{tab:geom}).

\begin{table*}[h!tbp]
 \renewcommand{\arraystretch}{1.0}
  \centering
    \begin{tabular}{|l|c||c|c|}\hline
    Cluster shape    &  Line of sight  &  $\epsilon _{y}^{geom}$& 
                     $\epsilon _{S}^{geom}$ \\ 
                     &   (in $r_c$)    & (in \%)                 &
                      (in \%)                \\ \hline\hline 
 {\em prolate}   &(1,0)&  0.9    &-17.7 \\ \hline
   &(1,1)&7.4 &4.0  \\ \hline
   &(0,1)&13.5 & 21.7 \\ \hline
 {\em oblate}   &(1,0)& -15.1     &-26.0 \\ \hline
   &(1,1)&-5.0 & 4.2 \\ \hline
   &(0,1)&0.9 & 19.5 \\ \hline

    \end{tabular}
    \caption{\small Relative errors on the Compton $y$-parameter and the surface 
brightness are shown. The prolate or oblate 
ellipsoid is 
supposed to have an axes ratio of $\zeta_1/\zeta_3=1.33$. Negative numbers 
indicate 
underestimations, whereas positive ones overestimations.}
    \label{tab:geom}
\end{table*}

\begin{figure}[h]
\begin{center}
\epsfig{file=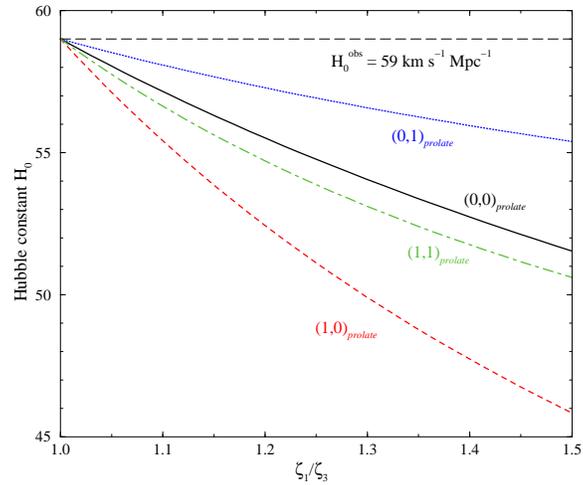,width=6.5cm,angle=-90}
\caption{\small The Hubble constant for an axisymmetric ellipsoidal 
shape (assuming a prolate, isothermal profile with infinite extension) for 
three different lines of sight parametrised in units of $r_c$. 
$H_o^{obs}$ is derived from the data of Mauskopf 
et al. (2000).}
\label{fig:obs3}
\end{center}
\end{figure}

The effect on the Hubble constant is shown in Figure \ref{fig:obs3} for 
a cigar shaped (i.e. prolate) cluster. We consider four different lines of sight, 
$(r_x, r_z)$=(1,0); (1,1); (0,0) and (0,1), given in units of the core 
radius $r_c$. For a strong flattening (i.e. $>1.4$) the value of the Hubble constant gets substantially modified.

\clearpage

\section{Outlook}
\label{SZ:S4}

In addition to our modifications, it should be noted that the commonly used expression for the fractional temperature decrement $\Delta T_{SZ}$ of the cosmic microwave background in clusters is based on the Kompaneets equation, which is derived under the assumption of non-relativistic electrons. However, the existence of many high-temperature galaxy clusters led to the need of taking into account the relativistic corrections for the electrons (Rephaeli 1995, Rephaeli \& Yankovitch 1997). Nozawa et al. (2000) presented useful fitting formulae for these relativistic corrections based on the calculations of Itoh et al. (1998). \\
MITO (Millimeter and Infrared Testa grigia Observatory), a 2.6 m ground 
based telescope (De Petris et al. 1996), is currently dedicated to 
cosmological observations in particular to the Sunyaev-Zel'dovich effect. 
As the first step, large and nearby clusters (diameter $\geq$ 5 arcminutes) 
have been selected; for example  in a recent paper D'Alba et al. (2000) 
have reported on the observations done on COMA cluster (Abell 1656, 
$z=0.0235$) and their preliminary results.\\
Therefore, in the analysis of coming data, it will be essential to take into account the relativistic corrections for high-temperature clusters and the 
possible effects, due to finite extension, polytropic index and geometry, that we have discussed above.

\section*{Acknowledgements}
The authors acknowledge J. Bartlett for the invitation to this workshop and the hospitality at 
the observatory of Toulouse (France). 
This work has been supported by the {\it D$^r$ Tomalla 
Foundation} and by the Swiss National Science Foundation. 
\section*{References}
\footnotesize{
Allen S.W., Fabian A.C., Johnstone D.A. et al. 1993, MNRAS 262, 901
\vskip2mm
\noindent
Cavaliere, A. \& Fusco-Femiano, R. 1976, A\&A 49, 137
\vskip2mm
\noindent
Cooray A. 1998, A\&A 339, 623
\vskip2mm
\noindent
D'Alba L., Melchiorri A., De Petris M. et al. 2000, {\texttt{astro-ph/0010084}}
\vskip2mm
\noindent
De Petris M., Aquilini E., Canonico M.  et al. 1996 New Astr. 1, 121
\vskip2mm
\noindent
Fabricant D., Rybicki G., Gorenstein P. 1984, ApJ 286, 186
\vskip2mm
\noindent
Grego L., Carlstrom J., Joy M. et al. 2000, {\texttt{astro-ph/0003085}}
\vskip2mm
\noindent
Hughes J.P. \& Birkinshaw M. 1998, ApJ 501, 1
\vskip2mm
\noindent
Hughes J.P., Gorenstein P., Fabricant D. 1988, ApJ 329, 82
\vskip2mm
\noindent
Inagaki Y., Suginohara T., Suto Y. 1995, Publ. Astron. Soc. Japan 47, 411
\vskip2mm
\noindent
Itoh N., Kohyama Y., Nosawa S. 1998, ApJ 502, 7
\vskip2mm
\noindent
Mauskopf P., Ade P., Allen W. et al. 2000, ApJ 538, 505
\vskip2mm
\noindent
Neumann D. M. \& B\"ohringer H. 1997, MNRAS 289, 123
\vskip2mm
\noindent
Nozawa S., Itoh N., Kawana Y \& Kohyama Y. 2000, ApJ 536, 31
\vskip2mm
\noindent
Pierre M., Le Borgne J.F., Soucail G., Kneib J.P. 1996, A\&A 311, 413
\vskip2mm
\noindent
Puy D., Grenacher L., Jetzer Ph., Signore M. 2000 to appear in A\&A, 
{\texttt{astro-ph/0009114}}
\vskip2mm
\noindent
Rephaeli Y. 1995, ApJ 445, 33
\vskip2mm
\noindent
Rephaeli Y. \& Yankovitch D. 1997, ApJ 481, L55
\vskip2mm
\noindent
Sulkanen M. 1999, ApJ 522, 59
\vskip2mm
\noindent
Sunyaev R. \& Zel'dovich Y. 1972, Comments Astroph. Space Phys. 4, 173
}
\end{document}